\documentclass[12pt,preprint2]{emulateapj}
\usepackage{natbib,amssymb,amsmath,graphicx}
\usepackage{textcomp}
\usepackage{longtable}

\begin{document}

\title{photometric study and period analysis of the contact binary XZ Leonis}

\author{ChangQing Luo\altaffilmark{1}, XiaoBin.Zhang\altaffilmark{1}, 
 Licai Deng\altaffilmark{1}, Kun Wang\altaffilmark{1}, Yangping Luo\altaffilmark{1} }

\altaffiltext{1}{Key Laboratory of Optical Astronomy, National
Astronomical Observatories, Chinese Academy of Sciences, Beijing
100012, China; changqingluo@bao.ac.cn}

\begin{abstract}
We present multi-color CCD photometry of the neglected
contact binary XZ Leo. Completely covered VRI band light curves and four times of minimum light were obtained.
 Combining the photometric and previously published radial velocity data, a revised photometric analysis was carried out 
for the binary system by applying the Wilson-Devinney code. With a hot spot placed on the 
massive primary component near the neck region of the common envelope, the light curves were satisfactorily modeled. 
The photometric solution combined with the radial velocity solution reveals that XZ Leo
is an A-type contact binary with a degree of contact of 24($\pm1)\%$.
 The absolute parameters of the components were determined as 
 $M_1$ = 1.74($\pm$0.06)$M_\odot$, $M_2$ = 0.61($\pm$0.02)$M_\odot$, 
 $R_1$ = 1.69($\pm$0.01)$R_\odot$, $R_2$ = 1.07($\pm0.01$)$R_\odot$,
$L_1$ = 6.73($\pm0.08$) $L_\odot$, $L_2$ = 2.40($\pm$0.04)$L_\odot$. 
Based on all the available data, the long-term orbital period behavior 
of the system was investigated. It indicates that the binary system 
was undergoing continuous orbital period increase in the past three decades
 with a rate of $dP/dt = + 6.12 \times {10^{-8}} days~ yr^{-1}$,
 which suggests a probable mass transfer from the secondary to the primary component at a rate of $dM/dt= 3.92\times 10^{-8} M_\odot~ yr^{-1}$.
  The binary system is expected to evolve into the broken-contact stage in $1.56 \times 10^6$ years. 
  This could be evidence supporting the Thermal Relaxation Oscillation theory.
\end{abstract}

\keywords{binaries: close - binaries: eclipsing; evolution - stars: individual(XZ Leo)}

\section{INTRODUCTION}
The light variation of XZ Leo ($BD+ 17^{\circ} 2165$, GSC 01412-01030) was discovered 
by \cite{hoff34}. \cite{pric47} pointed out this system was a W UMa-type contact binary. 
The first photoelectric light curves of XZ Leo were obtained by \cite{hoff84} and were analyzed by \cite{niar94}.
They noted that the light curves presented asymmetries 
between the first and second light maxima and that the second maxima
was slightly displaced to phase 0.76. 
They suggested that there might be two hot spots on both  components located near the neck region
of the common envelope. Subsequently, Rucinski \& Lu (1999) obtained the radial-velocity curves of this system. 
The radial-velocity solution revealed a mass ratio of 0.348 ($\pm {0.029}$) for the binary system  based on mass - centered sinusoids.
About ten years later complete BVRI light curves were obtained by \cite{Lee06a}.
They claimed a blue third light and a hot spot near the neck  
 of the primary component in their model. The solutions suggested that the system is a deep contact binary (f=33.6$\%$) with a small difference in temperature  of 
 $\bigtriangleup (T_1-T_2)=126 K$. Combining with the radial-velocity curves  \cite{ruci99},
 they determined the absolute parameters of XZ Leo as follows:  
 $M_1=1.84 M_\odot$, $M_2=0.63 M_\odot$, $R_1=1.75 R_\odot$,  $R_2=1.10 R_\odot$, 
 $L_1=7.19 L_\odot$, $L_2=2.66 L_\odot$.
 They also investigated all minimum light times to find the period increase
at a rate of $dP/dt = + 8.20 \times {10^{-8}} days~ yr^{-1}$ for XZ Leo following  \cite{qian01a}.
They suggested that the continuous period increase of XZ Leo could be due to mass transfer
from the secondary to the primary component  
and indicate that this system was undergoing an orbital expanding stage of Thermal Relaxation Oscillation (TRO) cycles, as suggested by \cite{qian01b}.
Therefore XZ Leo is a potentially key example as predicted by TRO theory, 
but there have been very few analyses of this system although it has been known for many decades.
 
Therefore, we have carried out new photometric observations of the binary.
 We present the multi-color CCD light curves in V, R, and I bands and analyzed the light curves with the Wilson - Devinney (W-D) code. 
 The new photometric solutions of the system are derived and the absolute parameters 
are determined by combining spectroscopic with
photometric solutions. 
Moreover, the variations in the orbital period of XZ Leo are analyzed. Period investigations are
 important in close binary studies, because they can not only provide information about some important physical processes  
(mass transfer, magnetic activity cycles), but also can give us a hint of the
evolutionary stage of the binaries. To obtain an accurate period and period changes, long-time-span data sets
and the precise times of light minimum data are important and necessary.
Here, we collected all available original data of XZ Leo to obtain reliable results. 
Finally, based on the period changes and the photometric
solutions of XZ Leo, the geometric structure and
evolutionary stage of this system are discussed .

\section{PHOTOMETRIC OBSERVATIONS}

New CCD photometric observations of XZ Leo in V-, R-, and I- bands were made on three nights in 2014 (February 
24, 27 and March 9) with the 85 cm reflecting
telescope at Xinglong station of the National Astronomical
Observatory, Chinese Academy of Sciences (NAOC). A 1024 $\times$ 1024 PI  CCD camera and 
a standard  Johnson - Cousins - Bessel multicolor filter system were used during the observation \cite{zho09}. 
The effective field of view is $16.'5$ $\times$  $16.' 5$. 
A total of 2110 individual observations were obtained in the three band passes (735 in V, 732 in R and 643 in I), 
which covered a timespan of about 24 hr. 
The twilight sky flat , bias and dark frames were taken on each observing day,
which were processed with the standard routines of CCDPROC in the IRAF package at first. 
Then, the instrumental magnitudes of stars detected in the program field were obtained by PHOT
of the aperture photometry package of IRAF. 
A star near the variable star was chosen as the comparison star (TYC 1412-247-1), 
which has a brightness and color similar to that of the variable star. 
Another star in the same field of view was selected as the check star (BD+17 2163a, GSC 01412-00423). 
The differential magnitudes of these stars were extracted in each frame. 
The corresponding coordinates of XZ Leo, the comparison and check stars are shown in Table 1.
Complete light curves in the V, R, and I bands are displayed in Figure 1 with open circles, where phases were computed using a period of
0.487739 days. This period is reconfirmed with the new data in this paper.

As shown in Figure 1, the observational light curves with open circles are typical EW-type light curves where the depths of both
light minimum are nearly the same. The light curves do not show
the O'Connell effects (different heights of the
two light maximum) of unequal light levels at the quadratures beyond the limits of the the 
observational error of about $\pm 0.01$ Mag. 
But the maxima I (phase =0.25) and maxima II (phase =0.75) of the light curves are displaced to around phases 0.24 and 0.76 respectively, 
which are shown in Figure 2. The solid line is the maxima I at phase =0.25 and maxima II at phase =0.75 (upper panel)
and the dotted line are phases at 0.24 and 0.76 (lower panel). 
These very same displacements are also found by Niarchos (1993) in the Hoffmann (1984) light curve as well as in the Lee et al. (2006) light curves. 
That means this displacement of the maxima is a long-term effect probably indicative of a long-term asymmetry in the system.
In our observations, a total of 4 eclipses of the contact binary XZ Leo were covered,
including 2 primary and 2 secondary ones. By applying the K-W method \cite{kvw56}, the epochs of minimum light 
with the mean values of V, R and I filters were determined as given in Table 2.

\begin{table*}
  \caption{Coordinates of XZ Leo, the comparison, and the check stars.}
   \begin{center}
   \begin{tabular}{llcc}\hline
Target &Names    &$\alpha 2000$ &$\delta 2000$ \\
\hline\noalign{\smallskip}
Variable star(V) &XZ Leo & $10^h$ 02$^m$ 34$^s$.19  & +17$^0$ 02$'$ 47$''$.15\\
The comparison star(C) &TYC 1412-247-1 & 10$^h$ 02$^m$  07$^s$.11 &+16$^0$ 59$'$ 21$''$.28\\
The check star(K) &BD+17 2163a &  10$^h$ 02$^m$ 20$^s$.20 & +17$^0$ 04$'$ 12$''$.65 \\
\noalign{\smallskip}\hline
  \end{tabular}
  \end{center}
\end{table*}

\begin{table}
  \caption{New times of light minimum for XZ Leo.}
   \begin{center}
   \begin{tabular}{lccc}\hline
HJD &Error     &Type &Filter \\
\hline\noalign{\smallskip}
2456713.0162 & 0.0001 & I & $VRI$\\
2456713.2607 & 0.0001 & II & $VRI$\\
2456716.1867 & 0.0001 & II & $VRI$\\
2456726.1872 & 0.0002 & I & $VRI$\\
\noalign{\smallskip}\hline
  \end{tabular}
  \end{center}
\end{table}


\begin{figure}
\center
\includegraphics[scale=.4]{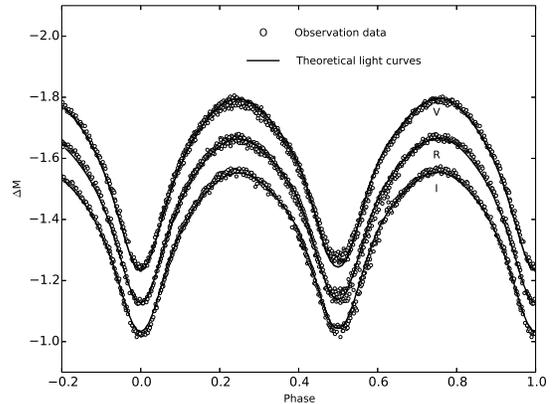}
\center
\caption{CCD photometric light curves in the V-, R-, and I- bands obtained using the
85 cm telescope at NAOC on 2014 February 24, 27 and March 9. 
 Open circles denote the observational data in V-, R-, and I- bands.  
 The solid line represents the theoretical light curves calculated using the W-D method discussed in Section 3.}
\center
\label{fdiff}
\end{figure}

\begin{figure}
\center
\includegraphics[scale=.4]{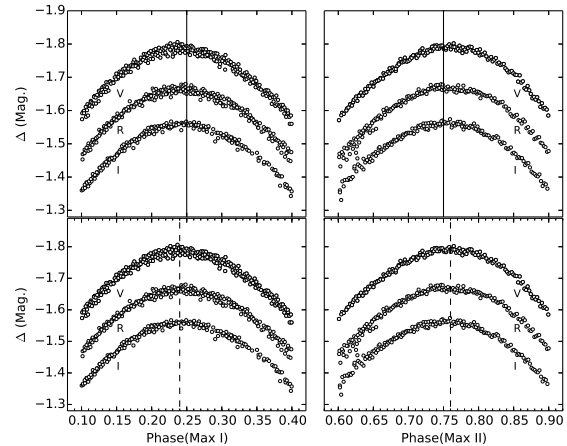}
\center
\caption{The Maxima I and Maxima II of the light curves. 
The solid lines in the top panels represent maxima I at phase = 0.25 and maxima
II at phase = 0.75. The dotted lines in the bottom panels show the displaced maxima at phases
0.24P and 0.76P, respectively.}
\center
\label{fdiff}
\end{figure}

\begin{figure}
\center
\includegraphics[scale=.4]{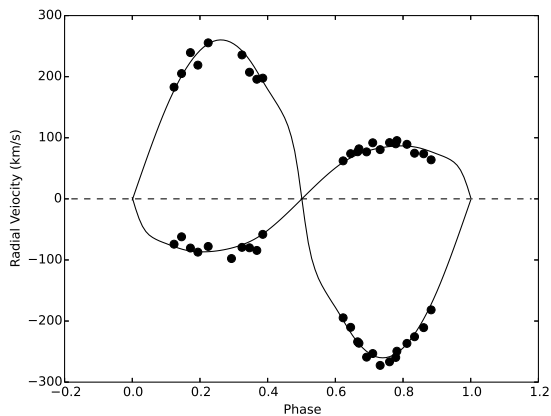}
\center
\caption{The Radial Velocity curves of XZ Leo. The solid circles are the observational data (Rucinsku \& Lu, 1999). The continuous
curves are the theoretical synthesis including proximity effects.}
\center
\label{fdiff}
\end{figure}

\section{ PHOTOMETRIC SOLUTIONS WITH THE W-D METHOD}
Combining the photometric and radial velocity data, the light curves in the V, R and I bands were analyzed using the W-D program
(Wilson $\&$ Devinney 1971 ; Wilson 1979, 1990, 1994, Wilson $\&$ Van Hamme
2003). 
We also assumed an effective temperature to be T1=7160 K for the primary component
(the star eclipsed at primary minimum), which was averaged from Harmanec's (1988) and Flower's (1996) tables corresponding to the spectral
 type A8V by the spectroscopy and the color index in the Hipparcos and Tycho Catalogues (Lee et al. 2006). 
 The gravity-darkening coefficients g1 = g2 = 1.0 (Lucy 1967) and the bolometric albedos A1 = A2 = 1.0 (Rucinski
1969) were used, because each component should have a radiative envelope or at most a shallow convective atmosphere(Lee et al. 2006).
 The initial mass ratio q was fixed to the spectroscopic mass ratio
q = 0.348($\pm0.029$) obtained by Rucinski \& Lu (1999).  A nonlinear limb-darkening law with a logarithmic form was applied in the light-curves synthesis.
The initial bolometric ($X_1$, $X_2$, $Y_1$, $Y_2$) and monochromatic limb-darkening coefficients ($x_1$, $x_2$, $y_1$, $y_2$,) of the components were taken from
Van Hamme (1993), which are listed in table 3. The adjustable parameters are as follows: mass ratio, q, 
the inclination, i, the mean temperature of star 2, T2, the monochromatic luminosity of star 1,
$L_{1V}$, $L_{1R}$, and $L_{1I}$, and the dimensionless potentials of star 1 and star 2 ($\varOmega _1$ = $\varOmega _2$ for mode 3).

As we have discussed in the above section that the light curves of XZ Leo show light maxima 
displacement. Initially, when no other physical processes are added into the binary model,  
we can not obtain a satisfactory fitting of the light curves. 
In order to solve this problem, a spot model was introduced.
Since the displacement of light maxima were noticed in all the light curves obtained by Hoffmann (1984), Lee et al. (2006) 
and  the present observations, it seems to be a long-lived phenomenon, which could not be caused by a cool spot related to
magnetic activity. Qian (2001a) and Lee et al. (2006) found that the orbital period of XZ Leo was continuously increasing,
suggesting probable mass transfer from the less-massive secondary to the primary component.
There are four spot parameters for each spot in the W-D program:
the spot temperature ($T_s$=$T_d$/$T_0$, $T_s$ is the ratio between the spot temperature $T_d$ and the photosphere surface
temperature $T_0$ of the star), the spot angular radius ($r_s$) in radians, the co-latitude of the spot center ($\theta$) in degrees
and longitude of the spot center ($\phi$) in degrees.
 The preliminary spot longitude could be found approximately using the phases of spot distortion in the light curves. 
 The other three parameters were calculated by fitting the theoretical light curves based on the observational data. 
 Finally, the best fitting of the photometric solutions were obtained with a hot spot.
 The observed data (marked with open circles) and the theoretical light curves (marked with solid lines) are shown in Figure 1.
 The results and the parameters of the hot spot
 are shown in Table 3.
 The corresponding geometric structure of the system with the hot spot is plotted in Figure 4. 
 The hot spot is located on the primary star near the neck region of the common envelope.

\begin{figure}
\center
\includegraphics[scale=.5]{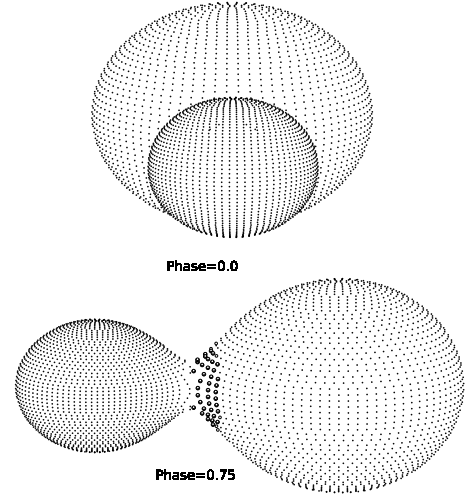}
\caption{Geometrical configuration and the spot distribution of XZ Leo.}
\center
\label{fdiff}
\end{figure}

\begin{table*}
  \caption{Photometric solutions of contact binary XZ Leo. }
   \begin{center}
   \begin{tabular}{llll}\hline
$~~~~~~~$Parameters &\underline{$~~~~~~$Best-fit Value(Model 1)$~~~~~~~$} &  \\
$~~~~~~~$           &Primary$~~~~~~~~~~~~~~~$Secondary &          \\  
\hline\noalign{\smallskip}
$~~~~~~~$ $g1=g2$  (deg)        & $~~~~~~~~~~~~~$  1.0&     \\
$~~~~~~~$ $A1=A2$  (deg)        & $~~~~~~~~~~~~~$ 1.0&     \\
$~~~~~~~$ $X_{bolo} $        & 0.642   $~~~~~~~~~~~~~~~~~~~$  0.641 &      \\
$~~~~~~~$ $Y_{bolo} $       & 0.259   $~~~~~~~~~~~~~~~~~~~$  0.253 &     \\
$~~~~~~~$ $x_{V}$           & 0.680 $~~~~~~~~~~~~~~~~~~~$    0.688  &   \\
$~~~~~~~$ $y_{V}$            & 0.297   $~~~~~~~~~~~~~~~~~~~$  0.290     &   \\
$~~~~~~~$ $x_{R}$            & 0.582   $~~~~~~~~~~~~~~~~~~~$ 0.592    &   \\
$~~~~~~~$ $y_{R}$            & 0.295   $~~~~~~~~~~~~~~~~~~~$  0.291  &       \\
$~~~~~~~$ $x_{I}$           & 0.490   $~~~~~~~~~~~~~~~~~~~$  0.501 &       \\
$~~~~~~~$ $y_{I}$           & 0.277   $~~~~~~~~~~~~~~~~~~~$ 0.275  &       \\
$~~~~~~~$ $i$  (deg)         & $~~~~~~~~~~~$ 77.89(12)  &\\
$~~~~~~~$ $q$=$M_{2}/M_{1}$ &   $~~~~~~~~~~~$ 0.346(16)  &   \\
$~~~~~~~$ $T$(K)            & 7160     $~~~~~~~~~~~~~~~~~~~$   6981(12)  & \\
$~~~~~~~$ $\Omega$          & 2.517(2)$~~~~~~~~~~~~~~~~$  2.517(2)&\\
$~~~~~~~$ $L_{1}/(L_{1}+L_{2})_{V}$  &   $~~~~~~~~~~~$ 0.748(1)  &    \\
$~~~~~~~$ $L_{1}/(L_{1}+L_{2})_{R}$  &   $~~~~~~~~~~~$ 0.738(2)  &  \\
$~~~~~~~$ $L_{1}/(L_{1}+L_{2})_{I}$   &   $~~~~~~~~~~~$ 0.734(2)  &   \\
$~~~~~~~$ $r$(pole) & 0.4541(4)$~~~~~~~~~~~~~$ 0.2821(4)&\\
$~~~~~~~$ $r$(side) & 0.4889(6)$~~~~~~~~~~~~~$ 0.2954(6)&\\
$~~~~~~~$ $r$(back) & 0.5185(7)$~~~~~~~~~~~~~~$0.3366(9)&\\
$~~~~~~~$ Latitude$_{spot}$(deg)  & $~~~~~~~~~~~$ 89.9(2) & \\
$~~~~~~~$ Longitude$_{spot}$(deg) &$~~~~~~~~~~~$  ~5.9(5)  & \\
$~~~~~~~$ Radius$_{spot}$(deg)    &$~~~~~~~~~~~$ 17.1(1) & \\
$~~~~~~~$ $T_{spot}/T_{2}$        & $~~~~~~~~~~~$ 1.079(5)  & \\
$~~~~~~~$ $f$(\%)        & $~~~~~~~~~~~$ 24(1)  &\\
\noalign{\smallskip}\hline
  \end{tabular}
  \end{center}
\end{table*}

\section{ORBITAL PERIOD VARIATIONS OF XZ LEO}
In order to investigate orbital period variations of XZ Leo, 
all available times of minimum light were collected from the available literature and
the database O-C gateway (http://var.astro.cz.ocgate/).

Adding the four times of  minimum light in this paper, 234 times of minimum light  were collected, 
in which 72 to visual, 35 to Photograph, 128 to photoelectric(Pe) and CCD. 
The Pe and CCD times of light minimum  for XZ Leo are listed in Table 4, 
where the second column shows the types of eclipses, 
 and the notation p and s refers to the primary and the secondary minima, respectively. The third column shows the observation method.  
The corresponding  Epoch and O-C values are listed in the forth and fifth columns based on the linear ephemeris
\begin{eqnarray}
Min.I &= 2456726.1706(7) + 0^d.48773679(4) \times E.
\end{eqnarray}
Then, together with all of the eclipse times data, the following quadratic ephemeris was derived by using the least-squares method:
\begin{eqnarray}
\begin{aligned}
Min.I =&2456726.1863(4) + 0^d.48773918(5) \times E\\ 
&+ 5.44 \times 10^{-11}  \times E^2.
\end{aligned}
\end{eqnarray}
The O-C fit curves for XZ Leo are plotted  in the upper panel of Figure 5 with solid lines. 
The observational data are plotted with open circles (Photoelectric and CCD data), triangles (Visual data)
and plus signs (Photographic data). Based on this ephemeris, a continuous period increase rate 
$dP/dt=+8.15 \times 10^{-8}$  $ days~ yr^{-1}$ was derived. 
This results is nearly as that of Lee et al.(2006), $dP/dt=+8.20 \times 10^{-8}$  $ days~ yr^{-1}$.
The residuals based on the ephemeris
(Equation (2)) are plotted in the lower panel of Figure 5, where no significant
systematic variation can be traced.
When considering only the photoelectric and CCD times of light minimum, 
we can obtain another quadratic ephemeris with the same method,
\begin{eqnarray}
\begin{aligned}
Min.I = &2456726.1851(2) + 0^d.48773886(5) \times E\\
&+ 4.08 \times 10^{-11} \times E^2. 
\end{aligned}
\end{eqnarray}
The continuous period increase can be clearly seen in the upper panel of Figure 6 (solid line). 
The period increase rate $dP/dt=+6.12 \times 10^{-8}$ $ days~ yr^{-1}$ was obtained, which is slightly smaller than that
derived from all of the minimum data (Equation 3). The resulting residuals based on Equation (4) 
are plotted in the bottom panel of Figure 6 where no  systematic deviations are evident. 
The orbital period's increase during
the contact phase can be interpreted as the conservative mass transfer
from the less massive component to the more massive one. The parameters of the system (e.g., the orbital
period, the mass ratio, the degree of contact) are changing
on thermal timescales of the components. The contact binary must undergo cycles around the state of marginal contact and
it will evolve into a broken-contact binary 
 at last as predictions of TRO theory (Lucy 1976 ; Flannery 1976 ; Robertson $\&$ Eggleton 1977).
 Therefore, XZ Leo should evolve to the broken-contact phase as suggested by Qian (2001b).
 Based on mass and angular momentum conservation and the equivalent radius of the Roche Lobe,
 we estimated the evolution time of XZ Leo from contact to broken-contact status.
 The equivalent radius of the Roche Lobe could be written as $R_{cr2}=r_{cr2}A$, and 
 \begin{eqnarray}
r_{cr2} = \frac{0.49q^{2/3}}{0.6q^{2/3}+1n(1+q^{1/3})}
\end{eqnarray}
 following Eggleton (1983), where q is the mass ratio (q = $M_2/M_1$) and A is the distance between two components. 
 We conclude that this contact binary will evolve to broken-contact phase in $1.56 \times 10^{6}$ yr.
 Corresponding 0.06 $M_\odot$ will be transfered from the secondary to primary component 
 and the mass ratio will be decreased to 0.304.

\begin{figure}
\center
\includegraphics[scale=.4]{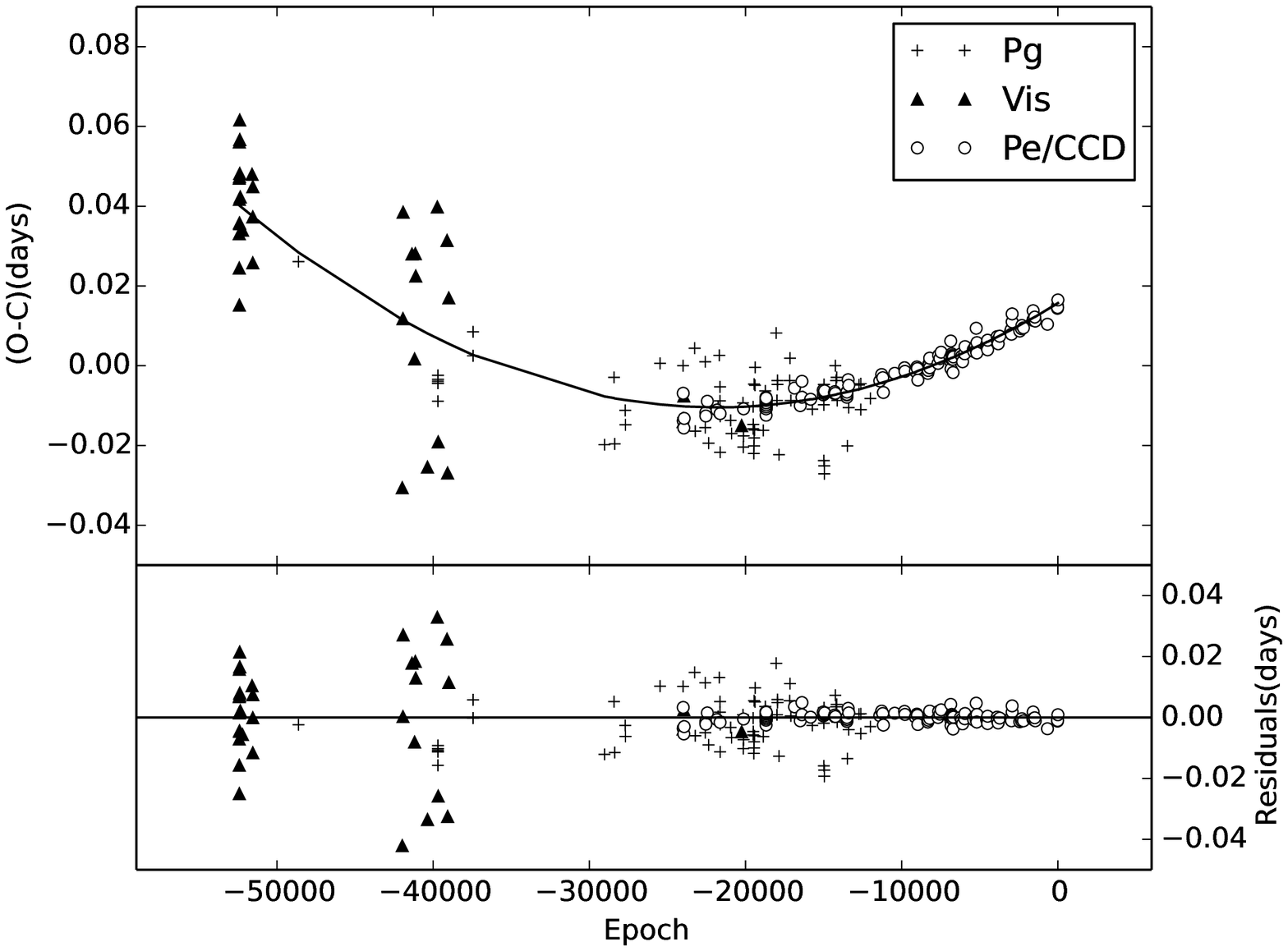}
\center
\caption{O-C curve of contact binary XZ Leo with all of the observation data. Upper panel: O-C
diagram computed with Equation (3) with the solid line, indicating that there is a long-term period increase. 
Bottom panel: resulting residuals. Open circles denote CCD and photoelectric data,
triangles denote the photographic data, plus signs the visual data}. 
\center
\label{fdiff}
\end{figure}

\begin{figure}
\center
\includegraphics[scale=.4]{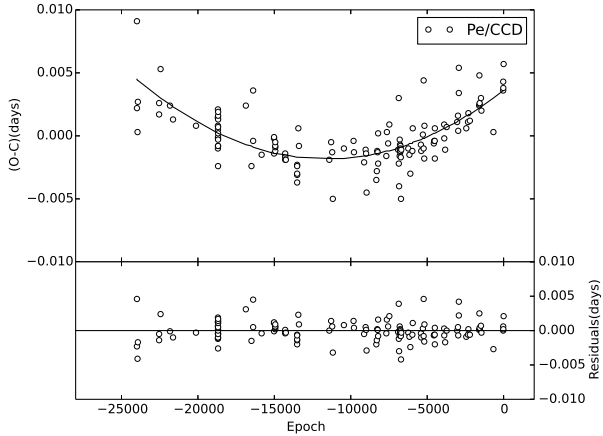}
\center
\caption{O-C curves of XZ Leo  with only photoelectric and CCD data.}
\center
\label{fdiff}
\end{figure}

\begin{table*}
  \caption{Photoelectric (Pe) and CCD Times of light minimum  for XZ Leo. }
   \begin{center}
   \begin{tabular}{cccccl}\hline
Epoch &Type &Method  &Epoch & O-C & References  \\
2400000+& & & & & \\
\hline\noalign{\smallskip}
45025.3580&	p	&Pe   &  -23990.0    &   0.0091 & \cite{hoff83}      \\
45025.5950&	s	&Pe   &  -23989.5    &   0.0022 & \cite{hoff83}      \\
45044.3710&	p	&Pe   &  -23951.0    &   0.0003 & \cite{Brau82}       \\
45055.3475&	s	&Pe   &  -23928.5    &   0.0027 & \cite{Brau82}       \\
45732.8154&	s	&Pe   &  -22539.5    &   0.0026 & O-C gateway$^{*}$         \\
45732.8145&	s	&Pe   &  -22539.5    &   0.0017 & \cite{Faul86}      \\
45779.3971&	p	&Pe   &  -22444.0    &   0.0053 & \cite{Hubs84}        \\
46079.8407&	p	&Pe   &  -21828.0    &   0.0024 & \cite{Faul86}      \\
46177.3872&	p	&Pe   &  -21628.0    &   0.0013 & \cite{Hubs85}      \\
46910.4568&	p	&Pe   &  -20125.0    &   0.0008 & \cite{Kesk89}       \\
47609.3834&	p	&Pe   &  -18692.0    &  -0.0010 & \cite{Wunder92}     \\
47609.3852&	p	&Pe   &  -18692.0    &   0.0008 & \cite{Hubs89}        \\
47609.3858&     p       &Pe   &  -18692.0    &   0.0014 & \cite{Hubs89}       \\
47609.3859&     p       &Pe   &  -18692.0    &   0.0015 & \cite{Hubs89}       \\
47609.3865&	p	&Pe   &  -18692.0    &   0.0021 & \cite{Hubs89}       \\
47612.3100&	p	&Pe   &  -18686.0    &  -0.0008 & \cite{Hubs89}       \\
47612.3106&     p       &Pe   &  -18686.0    &  -0.0002 & \cite{Hubs89}       \\
47612.3113&     p       &Pe   &  -18686.0    &  -0.0001 & \cite{Hubs89}       \\
47613.5278&	s	&Pe   &  -18686.0    &   0.0005 & \cite{Hubs89}       \\
47613.5294&	s	&Pe   &  -18683.5    &  -0.0024 & \cite{Hubs89}       \\
47614.5054&	s	&Pe   &  -18683.5    &  -0.0008 & \cite{Hubs89}       \\
47614.5059&     s       &Pe   &  -18681.5    &  -0.0003 & \cite{Hubs89}       \\
47614.5064&	s	&Pe   &  -18681.5    &   0.0002 & \cite{Hubs89}       \\
47616.4579&	s	&Pe   &  -18681.5    &   0.0007 & \cite{Hubs89}       \\
47616.4582&     s       &Pe   &  -18677.5    &   0.0013 & \cite{Hubs89}       \\
47616.4585&	s	&Pe   &  -18677.5    &   0.0016 & \cite{Hubs89}       \\
48500.4840&	p	&Pe   &  -18677.5    &   0.0019 & O-C gateway$^{*}$          \\
48680.4545&	p	&CCD  &  -16865.0    &   0.0024 &  \cite{Diet92}     \\
48733.3760&	s	&CCD  &  -16496.0    &  -0.0024 & \cite{Hubs92}      \\
48733.3800&	s	&CCD  &  -16387.5    &  -0.0004 & \cite{Hubs92}      \\
49004.5572&	s	&Pe   &  -16387.5    &   0.0036 & \cite{Hubs93}      \\
49400.3578&	p	&Pe   &  -15831.5    &  -0.0015 & \cite{Hubs94}      \\
49400.3579&     p       &Pe   &  -15020.0    &  -0.0002 & \cite{Hubs94}      \\
49401.3321&	p	&Pe   &  -15020.0    &  -0.0001 & \cite{Hubs94}      \\
49439.3759&	p	&CCD  &  -15018.0    &  -0.0014 & \cite{Hubs94}      \\
49439.3762&     p       &Pe   &  -14940.0    &  -0.0012 & \cite{Hubs94}      \\
49439.3763&     p       & Pe  &  -14940.0    &  -0.0009 & \cite{Hubs94}      \\
49439.3766&	p	&Pe   &  -14940.0    &  -0.0008 & \cite{Hubs94}      \\
49756.4048&	p	&Pe   &  -14940.0    &  -0.0005 & \cite{Ager96}       \\
49756.4050&     p       &Pe   &  -14290.0    &  -0.0019 & \cite{Ager96}       \\
49756.4051&     p       &Pe   &  -14290.0    &  -0.0017 & \cite{Ager96}       \\
49756.4053&	p	&Pe   &  -14290.0    &  -0.0016 & \cite{Ager96}        \\
49776.4021&	p	&Pe   &  -14290.0    &  -0.0014 & \cite{Diet95}      \\
50137.3277&	p	&Pe   &  -14249.0    &  -0.0019 & \cite{Ager96}       \\
50137.5702&	s	&Pe   &  -13509.0    &  -0.0023 & \cite{Ager96}       \\
50137.5708&     s       &Pe   &  -13508.5    &  -0.0037 & \cite{Ager96}       \\
50137.5709&     s       &Pe   &  -13508.5    &  -0.0031 & \cite{Ager96}       \\
50137.5715&	s	&Pe   &  -13508.5    &  -0.0030 & \cite{Ager96}        \\
50175.3742&	p	&CCD  &  -13508.5    &  -0.0024 & \cite{Ager97}       \\
50195.3700&	p	&CCD  &  -13431.0    &   0.0006 & O-C gateway$^{*}$        \\
51163.5287&	p	&CCD  &  -13390.0    &  -0.0008 & \cite{Ager99}       \\
51222.5464&	p	&CCD  &  -11405.0    &  -0.0019 & \cite{Ager00}       \\
51256.4434&	s	&CCD  &  -11284.0    &  -0.0005 & \cite{Ager00}       \\
51276.1930&	p	&CCD  &  -11214.5    &  -0.0013 & O-C gateway$^{*}$         \\
51629.8070&	p	&CCD  &  -11174.0    &  -0.0050 &  \cite{Nels01}      \\
51937.5703&	p	&CCD  &  -10449.0    &  -0.0010 & \cite{Ager02}       \\
51950.4944&	s	&CCD  &   -9818.0    &  -0.0004 & \cite{Ager02}       \\
52274.5538&     p       &CCD  &   -9791.5    &  -0.0013 & \cite{Csiz02}      \\
52274.5955&	p	&CCD  &   -9127.0    &  -0.0021 & O-C gateway$^{*}$         \\
52322.3948&	p	&Pe   &   -9029.0    &  -0.0011 & \cite{Ager03}       \\
52337.5143&	p	&CCD  &   -8998.0    &  -0.0014 & \cite{Ager02}       \\
52353.1189&	p	&CCD  &   -8966.0    &  -0.0045 & O-C gateway$^{*}$         \\
52664.2967&	p	&CCD  &   -8328.0    &  -0.0035 & O-C gateway$^{*}$         \\
52680.3927&	p	&Pe   &   -8295.0    &  -0.0028 & \cite{Hubs05}       \\
52692.5878&	p	&CCD  &   -8270.0    &  -0.0012 & \cite{Kotk06}        \\
52705.7566&	p	&CCD  &   -8243.0    &  -0.0013 &  \cite{Nels04}       \\

  \end{tabular}
  \end{center}
\end{table*}

\addtocounter{table}{-1}
\begin{table*}
  \caption{-continued}\label{Table 4}
   \begin{center}
   \begin{tabular}{cccccl}\hline
Epoch &Type &Method  &Epoch & O-C & References  \\
2400000+& & & & & \\
\hline\noalign{\smallskip}
52721.3633&	p	&Pe   &  -8211.0    &  -0.0022 &  \cite{Ager03}      \\
52721.6096&	s	&Pe   &  -8210.5    &   0.0002 &  \cite{Ager03}      \\
52983.2791&	p	&CCD  &  -7674.0    &  -0.0016 & O-C gateway$^{*}$         \\
53001.3274&	p	&CCD  &  -7637.0    &   0.0003 & O-C gateway$^{*}$          \\
53040.5894&	s	&CCD  &  -7556.5    &  -0.0006 &  \cite{Hubs05}       \\
53079.6099&	s	&CCD  &  -7476.5    &   0.0009 &  \cite{Hubs05}       \\
53359.8133&	p	&CCD  &  -6902.0    &  -0.0011 & O-C gateway$^{*}$          \\
53379.5708&	s	&CCD  &  -6861.5    &   0.0030 &  \cite{Alba05}        \\
53381.5166&	s	&CCD  &  -6857.5    &  -0.0022 & \cite{Hubs06}        \\
53390.7818&	s	&CCD  &  -6838.5    &  -0.0040 &  \cite{Oglo08}        \\
53409.5621&	p	&CCD  &  -6800.0    &  -0.0016 &  \cite{Hub05b}       \\
53410.5389&	p	&CCD  &  -6798.0    &  -0.0003 &  \cite{Hub05b}       \\
53421.2687&	p	&CCD  &  -6776.0    &  -0.0007 &  \cite{kim06}        \\
53427.1211&     p       &CCD  &  -6764.0    &  -0.0012 & \cite{Lee06a}         \\
53428.0964&     p       &CCD  &  -6762.0    &  -0.0013 & \cite{Lee06a}         \\
53429.0720&     p       &CCD  &  -6760.0    &  -0.0012 & \cite{Lee06a}         \\
53432.9743&     p       &CCD  &  -6752.0    &  -0.0008 &  \cite{Lee06a} \\
53433.2180&     s       &CCD  &  -6751.5    &  -0.0010 &  \cite{Lee06a} \\    
53445.4107&	s	&CCD  &  -6726.5    &  -0.0017 & \cite{Hub05b} \\
53451.5041&	p	&CCD  &  -6714.0    &  -0.0050 & \cite{Hub05b}) \\
53461.0189&     s       &CCD  &  -6694.5    &  -0.0011 & \cite{Lee06a}  \\
53683.6714&	p	&CCD  &  -6238.0    &  -0.0010 &  \cite{Hubs06}  \\
53720.2512&	p	&CCD  &  -6163.0    &  -0.0015 & O-C gateway$^{*}$     \\
53749.5140&	p	&CCD  &  -6103.0    &  -0.0030 & \cite{Hubs07}\\
53814.3850&	p	&CCD  &  -5970.0    &  -0.0012 & \cite{Hubs06}      \\
53826.3363&	s	&CCD  &  -5945.5    &   0.0006 & \cite{Sena07}       \\
54086.5432&	p	&CCD  &  -5412.0    &  -0.0007 & \cite{Dogr09}       \\
54097.2729&	p	&CCD  &  -5390.0    &  -0.0012 & O-C gateway$^{*}$         \\
54149.4622&	p	&CCD  &  -5283.0    &   0.0001 &  \cite{Hubs09}   \\
54174.3357&	p	&CCD  &  -5232.0    &  -0.0010 & O-C gateway$^{*}$        \\
54174.5850&	s	&CCD  &  -5231.5    &   0.0044 & \cite{Dogr09}        \\
54199.4560&	s	&CCD  &  -5180.5    &   0.0008 & \cite{Dogr09}        \\
54193.3567&	p	&CCD  &  -5193.0    &  -0.0018 & O-C gateway$^{*}$          \\  
54499.4135&	s	&CCD  &  -4565.5    &  -0.0006 & \cite{Yilm09}          \\
54509.4123&	p	&CCD  &  -4545.0    &  -0.0004 & \cite{Hubs10}          \\
54527.7025&	s	&CCD  &  -4507.5    &  -0.0004 & \cite{Nels09}          \\
54531.3591&	p	&CCD  &  -4500.0    &  -0.0018 & \cite{Hubs09}     \\
54531.6054&	s	&CCD  &  -4499.5    &   0.0006 & \cite{Hubs09}    \\
54828.6380&	s	&CCD  &  -3890.5    &   0.0009 & O-C gateway$^{*}$        \\
54831.5634&	s	&CCD  &  -3884.5    &  -0.0002 & \cite{Hubs10}     \\
54860.8268&	s	&CCD  &  -3824.5    &  -0.0011 & \cite{Diet09}          \\
54908.3830&	p	&CCD  &  -3727.0    &   0.0007 & \cite{Hubs10}          \\
55243.7032&	s	&CCD  &  -3039.5    &   0.0011 & \cite{Diet10a}          \\
55269.3100&	p	&CCD  &  -2987.0    &   0.0016 & O-C gateway$^{*}$ \\
55277.1126&	p	&CCD  &  -2971.0    &   0.0004 & O-C gateway$^{*}$  \\
55297.3567&	s	&CCD  &  -2929.5    &   0.0034 & O-C gateway$^{*}$ \\
55297.3587&	s	&CCD  &  -2929.5    &   0.0054 & O-C gateway$^{*}$ \\
55534.6384&	p	&CCD  &  -2443.0    &   0.0006 & O-C gateway$^{*}$ \\
55589.2655&	p	&CCD  &  -2331.0    &   0.0011 & O-C gateway$^{*}$  \\
55601.4597&	p	&CCD  &  -2306.0    &   0.0018 & \cite{Aren11}          \\
55649.7451&	p	&CCD  &  -2207.0    &   0.0012 &\cite{Diet11}          \\
55945.8032&	p	&CCD  &  -1600.0    &   0.0024 & \cite{Nels13}           \\
55952.8778&	s	&CCD  &  -1585.5    &   0.0048 & \cite{Diet12}          \\
55959.2161&	s	&CCD  &  -1572.5    &   0.0025 & O-C gateway$^{*}$   \\
55961.1672&	s	&CCD  &  -1568.5    &   0.0026 & O-C gateway$^{*}$    \\
56018.7197&	s	&CCD  &  -1450.5    &   0.0020 & \cite {Diet12}   \\
56003.3569&	p	&CCD  &  -1482.0    &   0.0030 & O-C gateway$^{*}$     \\
56400.3729&	p	&CCD  &   -668.0    &   0.0003 & O-C gateway$^{*}$      \\
56713.0162&     p       &CCD  &    -27.0    &   0.0036 &  This Paper      \\
56713.2607&     s       &CCD  &    -26.5    &   0.0043 &  This Paper     \\
56716.1867&     s       &CCD  &    -20.5    &   0.0038 & This Paper       \\
56726.1872&     p       &CCD  &      0.0    &   0.0057 & This Paper        \\
\noalign{\smallskip}\hline
  \end{tabular}

  \end{center}   Note. O-C gateway $^{*}$: http://var.astro.cz/ocgate/.
\end{table*}

\section{Discussions and Conclusions}
We have presented VRI bands time-series CCD photometry of the
contact binary XZ Leo. The light curve of this system has a long-lived asymmetric phenomenon in the placement 
of the light maxima from Hoffmann (1984) to Lee et al. (2006) to this paper.
This long-lived phenomenon may be caused by  mass transfer through the neck region.
Moreover, this system has a continuous period increase (Qian 2001a; Lee et al. 2006), 
the mass transfer will be from  secondary component (less massive) to the primary component (more massive).
A hot spot could be caused by the gas streams from the secondary component 
that impact the inner hemisphere of the primary component. 
 Therefore, a hot spot was introduced in our theoretical light curve. 
Based on our new light curves of XZ Leo and the published radial velocity data (Rucinski \& Lu (1999)),
we carried out a detailed light curve analysis for this system using  the Wilson-Devinney (W-D) code. 
A good fit to the observations was obtained with binary mode 3 under a hot-spot on
the primary component near the neck region of the common envelope.
The photometric solutions reveal that XZ Leo is an A-type system with a degree of contact of f=24($\pm1)\%$ . 
The effective temperatures of the components show that it is a hot contact binary with a small difference temperature of $\Delta(T_1-T_2)=179$K. 
 By combining our photometric solutions with radial velocity curves derived
by Rucinski \& Lu (1999), the absolute parameters of XZ Leo were computed and the results are listed in Table 5.

\begin{table}
  \caption{Absolute parameters of XZ Leo.}
   \begin{center}
   \begin{tabular}{lcc}\hline\hline
Parameter &Primary     &Secondary \\
\hline\noalign{\smallskip}
M($M_\odot$) & 1.74(6) & 0.61(2)\\
R($R_\odot$) & 1.69(1)  & 1.07(1)\\
L($L_\odot$) & 6.73(8)  & 2.40(4)\\
\noalign{\smallskip}\hline
  \end{tabular}
  \end{center}
\end{table}

To investigate the period variations of XZ Leo, all available times of light minimum  were collected.
Finally, 234 times of light minimum  were selected, where 128 came from the photoelectric and CCD observation.
 By combining our new times of  minimum light with all available timings, the new O-C diagrams of XZ Leo were analyzed, 
which reveals that the period of XZ Leo is increasing continuously at a rate of $dP/dt = +8.15 \times 10^{-8} days~ yr^{-1}$ with the whole data.
To obtain a more accurate period change, we used only the photoelectric and CCD data with 
the same method. The continuous period increase of  $dP/dt$ was also  presented and the period increase rate of 
$dP/dt = +6.12\times 10^{-8} days~ yr^{-1}$ was obtained, which is slightly smaller than that derived from all existing data.
The continuous period increase can be explained by mass transfer from the secondary (the less
massive component) to the primary component (the more massive component), which is in agreement with the predictions of  thermal relaxation oscillation (TRO Theory).
The TRO theory suggests that the evolution of W UMa stars must undergo oscillations 
around the status of contact and non-contact. This indicates that the contact binary XZ Leo is in an expanding TRO state before broken-contact phase at present, 
as suggested by Qian (2001a). By considering conservative mass transfer, a calculation with the well-known equation,
\begin{eqnarray}
\dot{P}/P = - 3\dot{M_2}(1/M_1-1/M_2)
\end{eqnarray}
leads to a mass transfer rate of $dM_2/dt= 3.92\times 10^{-8} M_\odot~ yr^{-1}$. 
This will cause the mass of the secondary component and hence the mass ratio to decrease. Conclusively, the contact binary of XZ Leo will evolve to broken-contact state.
Based on mass and angular momentum conservations, the time of evolution from contact to  broken-contact state was estimated to be 
$1.56 \times 10^6$ yr.
About 0.06 $M_\odot$ will be transfered from the secondary to the primary component and the mass ratio q will
decrease from 0.348 to 0.304.

\section{Acknowledgements} This work is supported by the National Natural Science
Foundation of China (NSFC) and the NSFC/CAS Joint
Fund of Astronomy through grants 11373037, 11303021,U1231202, and
2013CB834900. L.D. would like to acknowledge partial 
supported by the National Key Basic Research Program of China
2014CB845703. The authors are grateful to the anonymous referee for the valuable comments.

\end{document}